# A Gridless Compressive Sensing Based Channel Estimation for Millimeter Wave Massive MIMO Systems from 1-Bit Measurements

Mahdi Eskandari, Hamidreza Bakhshi

*Abstract*— This paper considers the problem of estimating the sparse millimeter wave (mmWave) massive multiple-input multiple-output (MIMO) OFDM channel from 1-bit quantized measurements. Unlike previous quantized on-grid approaches to compressive sensing, we introduce an off-grid convex scheme which uses binary atomic norm minimization (BiANM) to estimate sparse channel form 1-bit measurements. Moreover, in this paper, we propose a decoupled angle-delay version of BiANM called decoupled binary atomic norm minimization (DeBiANM) in order to reduce complexity. Further, for improving the accuracy of the estimation, we develop reweighted version of BiANM named reweighted binary atomic norm minimization (ReBiANM) that is a trade-off between atomic $\ell_0$ norm and atomic $\ell_1$ norm. Also, for improvement the performance of the DeBiANM, we introduce the reweighted version of DeBiANM, termed as reweighted decoupled binary atomic minimization (ReDeBiANM). Finally, through the simulation results, the performance of the proposed methods are studied.

I. INTRODUCTION

Recently, the growing demand for high data rate transmission, which is one of the main requirements of fifth generation (5G) wireless communication systems, has attracted considerable interests in both science and industry [1]. High operating frequency in mmWave systems has made the sampling rate of the analog-to-digital converters (ADCs) increase [2]. For example, at rates above 100 mega samples per second (MSPS), ADC power consumption increases quadratically in accordance with sampling frequency [2]. Therefore, using high-precision ADCs for mmWave systems are unavailable or expensive for portable devices [3]. A 1-bit ADC is essentially considered as one of the simplest devices used for quantization of an analog signal into a digital one. Since the number of bits needed to represent all quantization levels significantly affects the amount of ADC power consumption, 1-bit ADCs can be considered as the least power-hungry quantizers [4]. Because of the quantization nonlinearity, channel estimation with 1-bit ADCs is a challenge. Channel estimation in MIMO systems is even more difficult because the quantization applies for the linear combination of transmitting signals.

There exist several papers which have focused on mmWave MIMO channel estimation with few-bit ADCs. In [5], a least square (LS) approach is used to estimate the MIMO channel. Since they modelled the quantization error by the additive white Gaussian noise, the proposed channel estimation suffers from a significant estimation error. In [6], an expectation maximization (EM) is used for channel estimation. EM has high computational complexity because of needing matrix inversion and so many iterations to coverage. Approximate message passing (AMP) method is used for channel estimation in several recent works such as [7] and [8]. In these works, the authors have assumed that the channel coefficients follow a Gaussian independent and identically distributed (IID) distribution and sparse nature of the mmWave channels is not considered. An AMP-based channel estimation method for wideband MIMO channels with few-bit ADCs is investigated in [9]. In [10], the channel is estimated using an adaptive approach in a mmWave system with a 1-bit ADC. Atomic norm minimization (ANM) [11] has been applied to many applications such as frequency estimation [12], AoA estimation [13], and linear system identification [14]. In several papers such as [15] and [16], ANM has been used for MIMO channel estimation. An ANM approach for 1-bit quantized mmWave MIMO systems with hybrid architecture is introduced in [17]. Because of high computational complexity of the ANM approach, some recent works such as [18] and [19] have focused on low complexity channel estimation.

In this paper, we consider the problem of estimating a sparse mmWave channel of quantized 1-bit measurements in the noiseless case. Studying the atomic norm field, we have proposed a new off-grid 1-bit compressive sensing approach for channel estimation, called binary atomic norm minimization (BiANM). Also, for solving the BiANM with lower complexity, we propose decoupled binary atomic norm minimization (DeBiANM) approach. The atomic `0 norm directly exploits sparsity to the greatest extent but is NP hard to compute. To the contrary, as a convex relaxation, the atomic `1 norm can be efficiently computed [20]. Reweighted atomic norm minimization (ReANM) [20] fills the gap between two norms and sparsity will enhanced. At the rest of the paper, we developed the binary version of ReANM named reweighted binary atomic norm minimization (ReBiANM). Lastly, the reweighted version of DeBiANM termed as reweighted decoupled atomic norm minimization (ReDeBiANM) is proposed to enhance the performance of DeBiANM. Since, in the context of mmWave communications, we are dealing with noise, in the simulation section, we study the effect of noise on the performance of the proposed method. Besides, we used oversampling to compensate with errors caused by the noise.

The paper is structured as follows. The system and channel model of considered mmWave system is presented in section II. In Section III, we proposed the formulations of mmWave channel estimator. Simulation results are illustrated in Section IV. Finally, in Section V, the paper is concluded.

*Notations*: $a$ is a scalar, $\mathbf{a}$ is a vector, $\mathbf{A}$ is a matrix, and $\mathcal{A}$ represents a set. For a vector or matrix, the transpose and Hermitian operations are denoted by $(.)^T$ and $(.)^H$, respectively. $\mathbf{I}_N$ is the identity matrix of size $N \times N$. $\mathbf{A}^{-1}$ and $\text{trace}(\mathbf{A})$ are respectively the inverse and the trace of $\mathbf{A}$

vec(.) converts a matrix into a vector. $\mathbf{A} \otimes \mathbf{B}$ denotes Kronecker product of $\mathbf{A}$ and $\mathbf{B}$. $\mathbf{A} \succeq 0$ means that $\mathbf{A}$ is positive semidefinite (PSD). sign(.) is the signum function applied component wise to the real and imaginary part of the input argument. So the output of the signum function is one of the elements of the set $\{\pm 1 \pm j\}$. inf{.} is the infimum of the input set. Real and imaginary part of a complex number is denoted by $\Re\{.\}$ and $\Im\{.\}$, respectively. $\mathbb{E}[.]$ denotes expectation. Finally, $\mathbf{a}_K(\beta) = [e^{-j 2\pi \beta K'}]^T$, $K' = 0, 1, ..., K-1$ is the complex exponential sequence of length $K$ with $\beta \in [0,1]$.

## II. SYSTEM AND CHANNEL DESCRIPTION

We consider a point-to-point mmWave massive MIMO system. A single-antenna user equipment (UE) is communicating with a base station (BS) with a uniform linear array (ULA) of $M$ ($M \gg 1$) antennas. OFDM modulation with $N$ subcarriers is adopted for data transmission. The $M \times N$ complex baseband channel with the $(m,n)$-th element correspond to the complex channel gain of $m$-th antenna and $n$-th subcarrier is as follows [21].

$$\mathbf{H} = \sum_{l=0}^{L-1} \alpha_l \mathbf{a}_M(\theta_l) \mathbf{a}_N^H(\tau_l) = \sum_{l=0}^{L-1} \alpha_l \mathbf{A}(\Omega_l), \quad (1)$$

where $L \ll \min(M, N)$ is the number of propagation paths, $\alpha_l$ is the gain of the $l$-th path. $\theta_l \in [0,1]$ and $\tau_l \in [0,1]$ is the angle of arrival (AoA) and delay of the $l$-th path, respectively. There is no *a priori* knowledge about the statistic of the channel parameters. Besides, we assume that the two paths do not have a unique pair of path delay and AoA. The base station knows the value of $L$. The channel path gains follow i.i.d. complex Gaussian distribution of zero mean and variance $1/L$. Therefore, average received power per subcarrier will be $\sum_{l=0}^{L-1} \mathbb{E}|\alpha_l|^2 = 1$. Also, $\mathbf{A}(\Omega) = \mathbf{a}_M(\theta)\mathbf{a}_N^H(\tau)$. With this assumption, the received baseband signal is

$$\mathbf{Y} = \mathbf{H}\mathbf{X} + \mathbf{N}, \quad (2)$$

where $\mathbf{X} = \text{diag}(\mathbf{x})$ and $\mathbf{x} \in \mathbb{C}^{N \times 1}$ is the transmitted pilot symbols. $\mathbf{N} \in \mathbb{C}^{M \times N}$ is a noise matrix of i.i.d. complex Gaussian elements of zero mean and variance $\sigma^2$. With 1-bit ADC, the receiver will obtain

$$\mathbf{R} = \text{sign}(\mathbf{Y}) = \text{sign}(\mathbf{H}\mathbf{X} + \mathbf{N}), \quad (3)$$

the vectorized received signal is

$$\begin{aligned}\mathbf{r} &= \text{vec}(\mathbf{R}) \\ &= \text{sign}(\text{vec}(\mathbf{H}\mathbf{X}) + \text{vec}(\mathbf{N})) \\ &\overset{(a)}{=} \text{sign}((\mathbf{X}^T \otimes \mathbf{I})\mathbf{h} + \mathbf{n}) \\ &= \text{sign}(\mathbf{\Phi}\mathbf{h} + \mathbf{n}),\end{aligned} \quad (4)$$

where $\mathbf{h} = \text{vec}(\mathbf{H}) = \sum_{l=0}^{L-1} \alpha_l \underbrace{\mathbf{a}_N^H(\tau_l) \otimes \mathbf{a}_M(\theta_l)}_{\mathbf{a}(\omega_l)}$ and (a) results from the equality $\text{vec}(\mathbf{A}\mathbf{B}\mathbf{C}) = (\mathbf{C}^T \otimes \mathbf{A})\text{vec}(\mathbf{B})$. The problem in (4) can be analyzed with 1-bit compressed sensing framework

$$\overline{\mathbf{R}} = \text{sign}(\mathbf{A}\mathbf{k}), \quad (5)$$

where $\mathbf{k}$ is the sparse vector to be estimated via the measurement vector $\mathbf{A}$. In the channel estimation problem the matrix $\mathbf{A}$ is $\mathbf{\Phi} = (\mathbf{X}^T \otimes \mathbf{I}_M)$ and $\mathbf{k} = \mathbf{h}$. Note that with this binary measure, no information about the magnitude of the channel can be obtained. Therefore, the best case to do is recovering the normalized version of the channel. i.e., $\|\mathbf{h}\|_2 = 1$. According to the sparsity of the channel, we can use the gridless compressed sensing [22] to estimate the channel with the atomic norm minimization concept.

## III. GRIDLESS CHANNEL ESTIMATION BASED ON BINARY ATOMIC NORM MINIMIZATION

The constructing atoms of a signal is denoted by $\mathcal{A}$ and defines as follows.

$$\mathcal{A} = \{\mathbf{a}(\omega) \mid \omega \in [0,1] \times [0,1]\}. \quad (6)$$

The atomic $\ell_0$ norm $\|.\|_{\mathcal{A},0}$ of the channel is defend in the following [23], [24]

$$\|\mathbf{h}\|_{\mathcal{A},0} = \inf\{L \mid \mathbf{h} = \sum_{l=0}^{L-1} \alpha_l \mathbf{a}(\omega_l)\}. \quad (7)$$

The $\ell_0$ norm exploits sparsity to the greatest extent possible, but it is not convex and NP-hard to compute and cannot be globally solved with a practical algorithm. The atomic norm $\|.\|_{\mathcal{A}}$ is defined by its unit ball with the convex hull of $\mathcal{A}$ and is the convex relaxation of the atomic $\ell_0$ norm [23], [24]

$$\begin{aligned}\|\mathbf{h}\|_{\mathcal{A}} &= \inf\{\epsilon > 0 \mid \mathbf{h} \in \epsilon\text{conv}(\mathcal{A})\} \\ &= \inf\{\sum_{l=0}^{L}|\alpha_l| \mid \mathbf{h} = \sum_{l=0}^{L} \alpha_l \mathbf{a}(\omega_l), \mathbf{a} \in \mathcal{A}\}.\end{aligned} \quad (8)$$

With the above definitions, the estimation of the real-valued channel can be obtained with BiANM

$$\begin{aligned}\hat{\mathbf{h}}_{\text{BiANM}} &= \arg\min \|\mathbf{h}\|_{\mathcal{A}} \\ \text{s.t.} \quad &\text{sign}(\mathbf{\Phi}\mathbf{h}) = \text{sign}(\mathbf{y}) \\ &\|\mathbf{h}\|_2 = 1.\end{aligned} \quad (9)$$

The optimization problem in (9) has a convex objective function and non-convex constraints. In [25], a convex optimization problem for recovery of sparse signals from 1-bit measurements was introduced

$$\hat{\mathbf{h}}_{\text{BiANM}} = \arg\min \|\mathbf{h}\|_{\mathcal{A}}$$
$$\text{s.t.} \quad \mathbf{R}'\mathbf{\Phi h} \geq 0 \quad (10)$$
$$\|\mathbf{\Phi h}\|_1 = p,$$

where $p$ is any positive number, $\mathbf{R}' = \text{diag}(\mathbf{r})$, and the inequality is applied element wise. The first constraint is for consistency with 1-bit measurement. If there exists a vector $\mathbf{h}$ which satisfies the first constraint, then $\alpha \mathbf{h}$ for all $\alpha > 0$ will satisfy the first constraint too. Hence, only using the first constraint to solve the minimization problem for atomic $\ell_1$ norm will result the zero solution. In this case, the second constraint is used to prevent the zero solution. According to the first constraint $\|\mathbf{\Phi h}\|_1 = \langle \mathbf{r}, \mathbf{\Phi h} \rangle = p > 0$. Therefore, the second constraint will be equal to $\|\mathbf{\Phi h}\|_1 = \langle \mathbf{\Phi}^T \mathbf{r}, \mathbf{h} \rangle = p > 0$. This confirms that both objective function and constraints are convex. Comparing (9) and (10), both constraints $\|\mathbf{h}\|_2 = 1$ and $\|\mathbf{\Phi h}\|_1 = p$ prevent a non-trivial solution. Note that the above discussions can easily extend to the complex-valued case because the signum function applies separately to the real and imaginary part of input argument.

As considered in [11], the atomic norm is computed by semidefinite programming (SDP)

$$\|\mathbf{h}\|_{\mathcal{A}} = \arg\min \frac{1}{2MN}\text{trace}(\mathbf{T}_{2D}(\mathbf{u})) + \frac{\delta}{2}$$
$$\text{s.t.} \begin{bmatrix} \mathbf{T}_{2D}(\mathbf{u}) & \mathbf{h} \\ \mathbf{h}^H & \delta \end{bmatrix} \succeq 0, \quad (11)$$

where $\delta \in \mathbb{R}$, $\mathbf{u} \in \mathbb{C}^{MN \times 1}$ and $\mathbf{T}_{2D}(\mathbf{u})$ defined by its first row $\mathbf{u}$ of length $MN$, denotes a two-level Hermitian Toeplitz matrix constructed from the two-level Vandermonde structure of $\mathbf{A}(\mathbf{\Omega})$ [26]. With this assumption, (10) can be reformulated as

$$\hat{\mathbf{h}}_{\text{BiANM}} = \arg\min \frac{1}{2MN}\text{trace}(\mathbf{T}_{2D}(\mathbf{u})) + \frac{\delta}{2}$$
$$\begin{bmatrix} \mathbf{T}_{2D}(\mathbf{u}) & \mathbf{h} \\ \mathbf{h}^H & \delta \end{bmatrix} \succeq 0 \quad (12)$$
$$\text{s.t.} \quad \Re(\mathbf{R}')\Re(\mathbf{\Phi h}) \geq 0$$
$$\Im(\mathbf{R}')\Im(\mathbf{\Phi h}) \geq 0$$
$$\|\Re(\mathbf{\Phi h})\|_1 + \|\Im(\mathbf{\Phi h})\|_1 = 1.$$

When (12) is solved, the optimal value of $\hat{\mathbf{h}}_{\text{BiANM}}$ which is an estimate of $\mathbf{h}$ will be obtained.

As mentioned earlier, the optimization problem in (7), however, promotes sparsity to the greatest extent possible, it is non-convex and NP-hard to solve. But for using the sparsity nature of the channel to the great extent, an algorithm should be developed. Reweighted atomic norm minimization (ReANM) [20] is the solution. ReANM can mitigate the gap between two norms and leads to sparsity enhancement. The detailed implementation approach of ReANM algorithm is presented in [20]. The extension to the binary platform of ReANM is formulated as a binary reweighted atomic norm minimization (BiReANM) as follows

$$\hat{\mathbf{h}}_{\text{BiReANM}} = \arg\min \frac{1}{2MN}\text{trace}(\Theta_j \mathbf{T}_{2D}(\mathbf{u}))$$
$$+ \frac{\delta}{2}$$
$$\begin{bmatrix} \mathbf{T}_{2D}(\mathbf{u}) & \mathbf{h} \\ \mathbf{h}^H & \delta \end{bmatrix} \succeq 0 \quad (13)$$
$$\text{s.t.} \quad \Re(\mathbf{R}')\Re(\mathbf{\Phi h}) \geq 0$$
$$\Im(\mathbf{R}')\Im(\mathbf{\Phi h}) \geq 0$$
$$\|\Re(\mathbf{\Phi h})\|_1 + \|\Im(\mathbf{\Phi h})\|_1 = 1,$$

where $\Theta_j = (\mathbf{T}_{2D}(\mathbf{u}^{(j-1)}) + \zeta \mathbf{I}_{N_r N_t})^{-1}, j = 1, 2, ..., J$ and $\mathbf{u}^{(0)}$ equals $\mathbf{u}$ that obtains from (12) and $J$ is the number of iterations. $\zeta$ is a regularization parameter causes the optimization problem in (13) play $\ell_0$ norm minimization as $\zeta \to 0$ and $\ell_1$ norm minimization problem as $\zeta \to \infty$.

The main challenge of vectorization method is its considerable complexity. Because of vectorization in (13), the matrix size in the SDP constraint is $((MN+1) \times (MN+1))$ which leads to high complexity in both computation and memory as $M$ and $N$ become large. So a more efficient version of BiANM should be developed.

Recalling the channel model in (1) instead of vectorized atom set, one can adopt a new atom set $\mathcal{B}$ as

$$\mathcal{B} = \{\mathbf{A}(\Omega) \mid \Omega \in [0,1] \times [0,1]\}. \quad (14)$$

Our approach is to find the following atomic norm

$$\|\mathbf{H}\|_{\mathcal{B}} = \inf\{\sum_{l=1}^{L} |\alpha_l| \mid \mathbf{H} = \sum_{l=0}^{L} \alpha_l \mathbf{A}(\Omega), \mathbf{A}(\Omega) \in \mathcal{B}\}. \quad (15)$$

By reformulating (15) via SDP [13], we get the following result

$$\|\mathbf{H}\|_{\mathcal{B}} = \arg\min \frac{1}{2M}\operatorname{trace}(\mathbf{T}(\mathbf{u}_\theta))$$
$$+ \frac{1}{2N}\operatorname{trace}(\mathbf{T}(\mathbf{u}_\tau)) \quad (16)$$
$$\text{s.t.} \quad \begin{bmatrix} \mathbf{T}(\mathbf{u}_\theta) & \mathbf{H} \\ \mathbf{H}^H & \mathbf{T}(\mathbf{u}_\tau) \end{bmatrix} \succeq 0,$$

where $\mathbf{T}(\mathbf{u}_\theta)$ and $\mathbf{T}(\mathbf{u}_\tau)$ are one-level Hermitian Toeplitz matrices defined by the first row $\mathbf{u}_\theta$ and $\mathbf{u}_\tau$, respectively. So the final optimization problem for channel estimation with decoupled angle-delay point of view is as follows

$$\hat{\mathbf{H}}_{\text{DeBiANM}} = \arg\min \frac{1}{2M}\operatorname{trace}(\mathbf{T}(\mathbf{u}_\theta))$$
$$+ \frac{1}{2N}\operatorname{trace}(\mathbf{T}(\mathbf{u}_\tau))$$
$$\begin{bmatrix} \mathbf{T}(\mathbf{u}_\theta) & \mathbf{H} \\ \mathbf{H}^H & \mathbf{T}(\mathbf{u}_\tau) \end{bmatrix} \succeq 0 \quad (17)$$
$$\text{s.t.} \quad \Re(\mathbf{R})\Re(\mathbf{HX}) \geq 0$$
$$\Im(\mathbf{R})\Im(\mathbf{HX}) \geq 0$$
$$\|\Re(\mathbf{HX})\|_1 + \|\Im(\mathbf{HX})\|_1 = 1.$$

The main gain of the above problem formulation is its low complexity via decoupled view to the BiANM. Earlier in BiANM, the information of delay and angle of arrival of the channel was coded in a two-level block Toeplitz matrix which causes a lot of complexity in massive arrays. But here with the formulation based on Decoupled BiANM (DeBiANM) the delay and angle information are decoupled into two one-level Toeplitz matrices which decrease the SDP matrix constraint size to $(M+N) \times (M+N)$.

The minimum separation condition in the noiseless case that guarantees perfect channel recovery is as follows [26]

$$\min_{l \neq l'} \max\{|\theta_l - \theta_{l'}|, |\tau_l - \tau_{l'}|\} > \frac{1}{\min\{M,N\}}, \quad (18)$$

but, in decouple case the separation condition is [16]

$$\Delta_{\min,\theta} = \min_{l \neq l'}\{|\theta_l - \theta_{l'}|\} > d_1,$$
$$\Delta_{\min,\tau} = \min_{l \neq l'}\{|\tau_l - \tau_{l'}|\} > d_2. \quad (19)$$

The separation condition in the case of $M = N$ equals to $d_1 = d_2 = d = 1.19/[(N-1)/4]$ and yields

$$\min_{l \neq l'} \max\{|\theta_l - \theta_{l'}|, |\tau_l - \tau_{l'}|\} > 1.19/[(N-1)/4]. \quad (20)$$

The vectorized ANM is applicable in broader scenarios. The reason is that if any pair of two sources have enough separation in at least one dimension, $d_1$ and $d_2$ will be both less than $1.19/[(N-1)/4]$.

Similar to (13), we can use the ReANM idea to solve the optimization problem (17). We omit the proof under the page limit, but it can easily shown using [9] and [13] that we can use ReANM technique in DeANM to get better accuracy. As a result Reweighted decoupled binary atomic norm minimization (ReDeBiANM) optimization problem can be formulate as follows

$$\mathbf{H}_{\text{ReDeBiANM}} = \arg\min \frac{1}{M}\operatorname{trace}(\boldsymbol{\Theta}_j^{(\theta)}\mathbf{T}(\mathbf{u}_\theta))$$
$$+ \frac{1}{N}\operatorname{trace}(\boldsymbol{\Theta}_j^{(\tau)}\mathbf{T}(\mathbf{u}_\tau))\}$$
$$\begin{bmatrix} \mathbf{T}(\mathbf{u}_\theta) & \mathbf{H}^H \\ \mathbf{H} & \mathbf{T}(\mathbf{u}_\tau) \end{bmatrix} \succeq 0 \quad (21)$$
$$\text{s.t.} \quad \Re(\mathbf{R})\Re(\mathbf{HX}) \geq 0$$
$$\Im(\mathbf{R})\Im(\mathbf{HX}) \geq 0$$
$$\|\Re(\mathbf{HX})\|_1 + \|\Im(\mathbf{HX})\|_1 = 1,$$

where $\boldsymbol{\Theta}_j^{(\theta)} = (\mathbf{T}(\mathbf{u}_\theta^{(j-1)}) + \zeta \mathbf{I}_{MN})^{-1}$ and $\boldsymbol{\Theta}_j^{(\tau)} = (\mathbf{T}(\mathbf{u}_\tau^{(j-1)}) + \zeta \mathbf{I}_{MN})^{-1}, j = 1, 2, ..., J$. $\mathbf{v}_\theta^{(0)}$ and $\mathbf{u}_\theta^{(0)}$ equals to $\mathbf{u}_\theta$ and $\mathbf{u}_\tau$ that results from (17).

## IV. SIMULATION RESULTS

In this section, we evaluate the performance of the proposed mmWave channel estimators with 1-bit measurement. A BS with $M = 64$ antenna is assumed that is communicating with a single antenna UE with $N = M$ OFDM subcarriers. Without loss of generality, we assume that all the pilot symbols are equal to $1$. Path delays and AoAs were independently generated and uniformly distributed in $[0, 1/4]$ and $[0, 1]$, respectively. This range is a reasonable value in practical systems [27]. The signal-to-noise ratio (SNR) could be defined as $\text{SNR} = 1/\sigma^2$. The number of paths is set to be $L = 3$. For acquiring the results, Monte-Carlo simulation is used, which averaged over many independent trials. For the ReBiANM and ReDeBiANM algorithms, we initial the value of $\zeta$ equals to 1 and halve $\zeta$ when beginning a new iteration. The iteration number for both ReBiANM and ReDeBiANM is set to be $J = 5$. Lastly, CVX is used for solving the optimization problems.

Because the mmWave systems are joint with noisy measurements, we consider an oversampling technique to compensate for the possible errors in the measurements because of the high noise level. In this situation, five samples with different noise realizations are taken in the same measurement. The final selection is made after a majority vote. Our metric is the normalized mean square error (MSE), defines as $\mathbb{E}\left[\frac{\|\hat{\mathbf{h}} - \mathbf{h}\|^2}{\|\mathbf{h}\|^2}\right]$ where $\hat{\mathbf{h}}$ is the vectorized estimate of vectorized channel $\mathbf{h}$.

Fig. 1, compares gridless algorithms as a function of SNR. In the case of 1-bit ADC, it is clear that BiANM

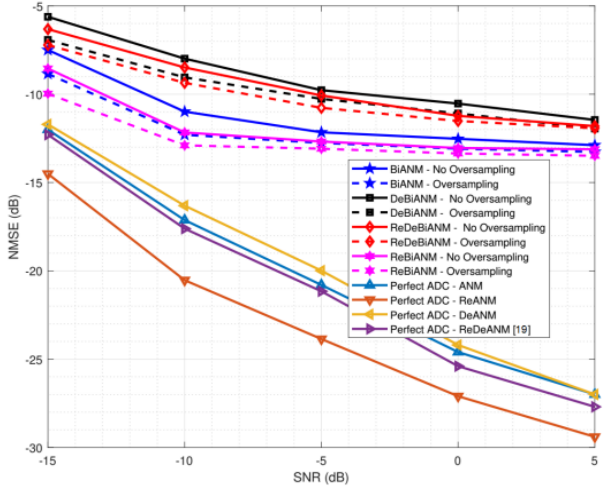

Fig. 1. Performance compression of different channel estimation schemes

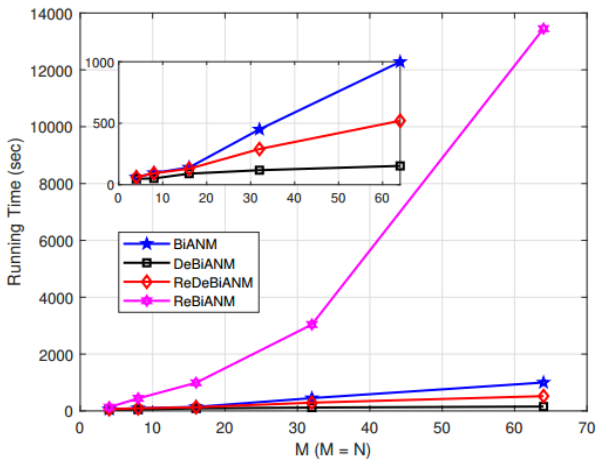

Fig. 2. Illustration the computational complexity with a metric of run time vs. number of transmitter antennas ($M$) and OFDM subcarriers ($N$), $N = M$

overcomes other schemes, but in contrast to perfect ADC case, Reweighted algorithms, i.e., ReBiANM and ReDeBiANM do not have much better performance than BiANM and DeBiANM, it is because the amplitude of the channel is almost lost and the accuracy of the channel estimation will decrease. Moreover, BiANM and ReBiANM have better performance than DeBiANM and ReDeBiANM, respectively. This is because of minimum separation condition. Also the oversampling methods act better than corresponding no oversampling case in low SNR regime. The reason is that in high SNR's the effect of noise is lower than low SNR's. So the NMSE in high SNR is similar to no oversampling case.

The running speeds of these proposed methods is plotted in Fig. 2. As it is obvious, the running time of DeBiANM is smaller than BiANM. Also the running speed of ReDeBiANM is very smaller than ReBiANM. Also proposed schemes exhibit a considerable benefit in computational efficiency for massive arrays.

## V. CONCLUSION

In this work, we proposed a method to estimate the channel from 1-bit measurements in a sparse and continuous manner. The channel estimation was formulated as a new version of ANM called BiANM. We provide DeBiANM algorithm to solve BiANM effectively for large-scale problems by introducing a new atom set. Moreover, we provide ReDeBiANM to solve DeBiANM for further improvement. Numerical experiments verified the accuracy of proposed methods.